\documentclass{IEEEcsmag}

\usepackage[colorlinks,urlcolor=blue,linkcolor=blue,citecolor=blue]{hyperref}

\usepackage{upmath}
\usepackage{xcolor}
\usepackage{booktabs}
\usepackage{graphicx}
\usepackage{wrapfig}

\jvol{XX}
\jnum{XX}
\paper{}
\jmonth{}
\jname{}
\pubyear{2022}

\usepackage{ifthen}
\usepackage{amssymb}
\DeclareOldFontCommand{\sf}{\normalfont\sffamily}{\mathsf}
\newboolean{showcomments}
\setboolean{showcomments}{true}
\ifthenelse{\boolean{showcomments}}
  {\newcommand{\nb}[2]{
    \fcolorbox{gray}{yellow}{\bfseries\sffamily\tiny#1}
    {\sf\small$\blacktriangleright$\textit{#2}$\blacktriangleleft$}
   }
   
  }
  {\newcommand{\nb}[2]{}
   
  }

\begin{document}


\title{Knights and Gold Stars: A Tale of InnerSource Incentivization}

\author{Tapajit Dey }
\affil{University of Limerick and Lero, the SFI Research Centre for Software}

\author{Willem Jiang}
\affil{Huawei}

\author{Brian Fitzgerald}
\affil{University of Limerick and Lero, the SFI Research Centre for Software}


\begin{abstract}
Given the success of the open source phenomenon, it is not surprising that many organizations are seeking to emulate this success by adopting open source practices internally in what is termed InnerSource. However, while open source development and InnerSource are similar in some aspects, they differ significantly on others, and thus need to be implemented and managed differently. To the best of our knowledge, there is no significant account of a successful InnerSource incentivization program. Here we describe a comprehensive InnerSource incentivization program which was implemented at Huawei. The program is based on theories of motivation, both intrinsic and extrinsic, and also includes incentives at the individual, project, and divisional level, which helps to overcome the barriers that arise when implementing InnerSource. The program has had very impressive early results, leading to significant increases in the number of InnerSource projects, contributors, departments, and lines of code contributed. 
\end{abstract}

\maketitle
\chapterinitial{Introduction}
Over the past 25 years, the Open Source Software (OSS) phenomenon has arguably been the greatest disruptive force in the software field, demonstrating that high quality software can be produced by a globally-distributed and largely-volunteer workforce, certainly countering early scepticism as to the future of OSS~\cite{fitzgerald2005has}. This has had major implications in both research and practice, and organizations around the globe are incorporating open source practices into the way they build and ship their own software. This software development approach, commonly known as ``InnerSource'', focuses on leveraging the Open Source development model for developing proprietary software. The popularity of InnerSource has increased dramatically over the past few years, with many organizations, including major technology companies like Microsoft, Huawei, Tencent, and PayPal, and even non-IT companies like Bloomberg, Bosch, and Nike embracing the culture. InnerSource Commons, an online community of InnerSource practitioners, has over 1,300 subscribers across 500 organizations.\footnote{Source: \url{https://innersourcecommons.org/community/action/}, Accessed on 2022-01-09.}

A number of significant benefits may arise from successful implementation of InnerSource \cite{capraro2016inner, InnerSourceAdopting}. These include:
\begin{itemize}
  \item \textit{Increased code reuse} - Sharing of code and collaborative development across organizational divisions ensures increased code reuse.
  \item \textit{Quality improvement} - Greater level of review of requirements, code, and testing facilitates bug discovery (by leveraging Linus's law --- ``Given enough eyeballs, all bugs are shallow'') and improves the overall quality of the software product.
  \item \textit{Open innovation} – Access to broad and deep knowledge beyond the confines of the department originally developing the software facilitates open innovation.
  \item \textit{Improved staff mobility} - Greater familiarity with development and interaction with colleagues outside departmental silos facilitates staff mobility across divisions.
  \item \textit{Attracting and recruiting talented employees} – Most fresh graduates nowadays are familiar with the open source model and would expect and be attracted to a development environment conducive to social coding rather than a tightly controlled traditional one. 
\end{itemize}

\noindent However, successful implementation of InnerSource has long been a challenge \cite{van2009commodification, riehle2016inner}, and significant institutional change may be required \cite{lindman2010open}. A number of barriers can arise in organizations which hinder InnerSource implementation, as identified by the `2020 State of InnerSource Survey'\footnote{State of InnerSource Survey, \url{https://tapjdey.github.io/InnerSource\_Survey\_2020}, Accessed on 2022-01-09.}.
These include:
\begin{itemize}
    \item \textit{Time constraints} - Developers in organizations are fully committed to their ``day job'' and often do not have the capacity to take on additional development tasks for InnerSource.
    \item \textit{Lack of recognition} – The traditional performance appraisal mechanisms are typically designed for silos of organizational departments and divisions, and often do not recognize the pan-organizational nature of InnerSource. Moreover, the developed code is not viewed externally, so there is limited visibility of InnerSource contributions outside the company, which doesn't help an individual's career as much as OSS contributions.
    \item \textit{Lack of community support} - The ``free rider'' phenomenon, whereby individuals take more than they contribute, which has been reported for open source, applies to InnerSource also. Contributors to InnerSource might want to grow through participation, but without the support of an active and supportive community, they might end up working on maintenance issues against their will, and might experience frustration and burn-out in the process.
    \item \textit{Lack of management support} – Lower and middle management often focus more on short term targets, prioritizing local development activities over InnerSource work since there is no immediate benefit to their individual divisions.
    \item \textit{Liable to fix code subsequently} – Developers may fear that contributing code to InnerSource will require them to maintain that code indefinitely, or be responsible for any issues that might arise through use of the code.
    \item \textit{Fear of contributions being ``hijacked'' without due credit} – Somewhat in opposition to the previous concern, developers may fear that others will take credit for their InnerSource contributions.
\end{itemize}
In order to successfully implement InnerSource at an organization and realize its benefits, it is essential to address the barriers to InnerSource adoption. This can be achieved through appropriate incentivization~\cite{incentives}, however, designing a suitable incentive program for facilitating InnerSource adoption is not trivial --- InnerSource lies at the intersection of open source and closed source development, thus the motivations for those involved are not exactly the same as either in isolation~\cite{aagerfalk2015software}. Moreover, to the best of our knowledge, there is no existing research on a structured incentive program tailored to facilitate InnerSource adoption. 

In this article, we address this gap by presenting the InnerSource implementation at Huawei and the structured incentive program proposed for facilitating its adoption.

\section{Motivations for Open Source Participation}

While InnerSource and open source share many similarities, they also differ on a number of dimensions, including the nature of the workforce and the locus of control \cite{aagerfalk2015software}. We are not aware of research on an incentive program designed specifically for InnerSource. However, a lot of research has focused on the motivations underpinning OSS participation more generally. These include the following broad categories:\cite{von2012carrots}
\begin{itemize}
  \item Intrinsic motivators (e.g., Ideology, Kinship, Enjoyment and Fun)
  \item Extrinsic Motivators (e.g., Pay, Career)
  \item Internalised Extrinsic (e.g., Reputation, Learning, Reciprocity, Own-use)
\end{itemize}
We briefly elaborate on each of these categories below.

Intrinsic motivation has long been identified as a key factor in open source. \textit{Ideology} was at the heart of the Free Software movement which pre-dated open source. Closely related to this is the concept of \textit{Kinship}. Helping a group to which open source developers feel a tie of kinship is thus a strong motivator. Also, the idea of \textit{Enjoyment and Fun} is intrinsically motivating, and has been identified as a core aspect of the ‘hacker culture’ in the 1980s that also fuelled the emergence of open source. 

Extrinsic motivation, such as \textit{pay} and \textit{career}, has become a major factor in open source, especially over the past few years. As more companies become involved in open source and as it becomes a key strategic component in the software industry, the model whereby developers are paid to work on open source becomes increasingly common\cite{riehle2014paid}. Also, open source contributions could be clearly attributed to individual developers and hence these could form part of a developer’s achievement that could support career progression, employment, and promotion.

Some motivators, which might appear at first glance to be extrinsic, can also be internalized by developers to the extent that they act as self-regulating in terms of behavior. For example, \textit{reputation} is a major factor underpinning developer involvement in open source, especially among peers. \textit{Learning} is also a significant motivator in that developers can improve their skills through the transparency of contributions which address technical challenges. \textit{Reciprocity} also belongs in this internalized extrinsic category in that developers are happy to contribute according to their strengths and expect their contributions to be reciprocated by others. \textit{Own-use value} follows from this as developers seek to consume open source products for their personal use.

\section{InnerSource Adoption at Huawei}


Huawei is a Chinese multinational technology corporation headquartered in Shenzhen, China. It designs, develops and sells telecommunications equipment, consumer electronics, and various smart devices. Huawei has almost 200,000 employees and has deployed its products and services in more than 170 countries. It was ranked as the second-largest R\&D investor in the world by the EU Joint Research Centre in 2021. 

Due to its size, there is a lot of human potential in Huawei that is hard to utilize fully in a traditional governance system. Therefore, Huawei decided to adopt InnerSource to promote pan-organizational knowledge-sharing and open innovation. There are four key mechanisms supporting the InnerSource adoption at Huawei, viz. the governance mechanism, the infrastructure support, the promotion \& training, and the structured incentive program for InnerSource contributors.
We discuss the first three mechanisms briefly below in order to provide some context before we discuss the incentive program in detail in the following sections.

\subsection{InnerSource Governance }\label{ss:IS-gov}

In order to oversee the task of InnerSource adoption, Huawei have established a very high-level \textit{InnerSource Foundation} across the entire organization. This is chaired by a Huawei Fellow, and its members include the President of each Product Line and the Director of each R\&D department. The mandate of this Foundation is to approve Huawei's InnerSource policies, charters, mechanisms, and plans. It is also charged with promoting and advocating the InnerSource concept throughout the organization and building a healthy InnerSource ecosystem, culture, and climate. 

This InnerSource Foundation is further underpinned by an InnerSource \textit{Technical Committee} (TC), again chaired by a Huawei Fellow and comprising as members the Chief Software Engineers of each Product Line. The TC is the primary committee coordinating InnerSource activities at Huawei. 
The remit of the Technical Committee is to:
 \begin{enumerate}
     \item Select and establish the appropriate InnerSource technology communities and projects based on the overall strategy and direction of the Board of Directors.
     \item Define the phases (preparation, incubation, and graduation) and requirements of InnerSource projects, review and approve InnerSource communities and projects, and guide the operation of each InnerSource project.
     \item Responsible for developing, maintaining and publishing the Huawei InnerSource License, and arbitrating technical disputes arising from non-compliance with InnerSource usage.
     \item Responsible for promoting and spreading the culture of InnerSource, and motivating and recognizing InnerSource technology communities, projects, and individuals with outstanding contributions.
 \end{enumerate}

In addition to the TC, there are \textit{Technical Community Committees} (TCCs) specific to each product line for overseeing InnerSource activities and facilitating communications among members of different product lines. The TCCs review and approve InnerSource communities and projects, appoint \textit{Project Management Committees} (PMCs) for InnerSource projects, identify InnerSource opportunities in the technology domain, build a healthy and vibrant InnerSource technology ecosystem, evangelize the InnerSource culture, motivate and commend outstanding InnerSource technology communities, projects, and individuals, and publicize InnerSource concepts to other departments.

The PMC is the technical management committee for each InnerSource project, consisting of project maintainer representatives and other stakeholders, including management. PMCs are tasked with ensuring three key requirements:
\begin{enumerate}
    \item The InnerSource project is consistent with the  strategy and values of the department. 
    \item Ensuring credit for contributions to the InnerSource project is attributed correctly to the the actual contributors, and also that contributions do not result in an unreasonable demand for maintenance subsequently.
    \item Ensuring that contributors have autonomy and can focus on building up personal reputation and influence through InnerSource.
\end{enumerate}

\subsection{InnerSource Infrastructure}

Infrastructure support is a necessary component for successful InnerSource implementation \cite{linaaker2014infrastructure, gurbani2010managing}. A team of engineers at Huawei help in the design and implementation of required components, e.g., an InnerSource portal and dashboard showcasing the active projects and individual profiles of InnerSource contributors, their activity and badges/awards. They also help in tracking the activity of projects and developers using various metrics as required by the wider community of InnerSource participants in the organization.

\subsection{InnerSource Promotion \& Training}

Since InnerSource operates quite differently from traditional development operations at Huawei, having standard procedures is key for ensuring the InnerSource community can thrive. The two main goals for setting up the processes are firstly spreading awareness about InnerSource  within the company to help attract new developers via various online and offline promotional activities, such as meet-and-greet sessions, hackathons, new developer meets, and secondly, ensuring smooth operation of existing InnerSource projects, e.g, asynchronous and collaborative development, meritocracy, etc. by providing training to InnerSource participants. The processes are overseen by the InnerSource governing bodies at different levels, who design and promote InnerSource standards and practices, ensure that the rules are followed, and act as arbiters in case of any disputes.

\section{Designing an InnerSource Incentive Program at Huawei}

In order to facilitate the adoption of InnerSource and recognize employee contributions to InnerSource projects, Huawei has designed and implemented a structured incentive program, which is a complement to Huawei's existing employee appraisal system, as discussed below. 

In designing the InnerSource incentive program at Huawei, a number of the fundamental concepts underpinning motivation were drawn on. These include Maslow's proposed hierarchy of needs from physiological and safety needs to those of esteem, and self-actualization. 
In a similar vein, Herzberg’s two-factor theory~\cite{herzberg1966work} identifies both hygiene factors and motivating factors. Hygiene factors, if not present, will cause an employee to work less and lead to dissatisfaction, while motivating factors will encourage an employee to work harder and can bring about satisfaction. More contemporary thinking on motivation was drawn from Daniel Pink's Drive concept which identifies three categories of motivation -- autonomy, mastery and purpose~\cite{pink2011drive}. 

These concepts were vital in designing the InnerSource incentive program at Huawei. Theories on general motivation were combined with empirical and contextual knowledge, viz. the research findings on OSS developer motivation and research on existing employee appraisal frameworks, as well as the management styles of large OSS communities such as Linux and the Apache Software Foundation, to come up with a concrete strategy for tackling the barriers to InnerSource adoption, and motivating individual developers, potential project maintainers, and management to adopt InnerSource.

\subsection{Hygiene/Deficiency Factors}

Many of the barriers to InnerSource adoption are conceptually similar to the lower levels of Maslow's hierarchy of needs, e.g. Physiological and Safety needs, and can be thought of as `hygiene' factors - they are necessary precursors to initiate InnerSource adoption but do not actively create an InnerSource culture. These factors include - time constraints, lack of support from management and peer groups, fear of not receiving due credit for InnerSource contributions, and having to maintain code indefinitely. 

Research on InnerSource suggests that support from top management, colloquially known as `Executive Air Cover', is essential for effective adoption of InnerSource~\cite{InnerSourceAdopting}. For Huawei, the InnerSource Foundation and the active PMCs ensure employees have adequate support from management for undertaking InnerSource work. This also reassures managers at lower levels of the organization, who control the workload of developers in their teams, that there is higher-level managerial support for InnerSource activities. The regular publicizing activities, including developer meet-ups and workshops, also educate  developers about how InnerSource development works (since it can be significantly different from the practices followed by individual teams). This in turn results in more realistic expectations about InnerSource and helps establish a more supportive environment for InnerSource.

\begin{figure*}[!htb]
    \centering
    \includegraphics[width=\textwidth]{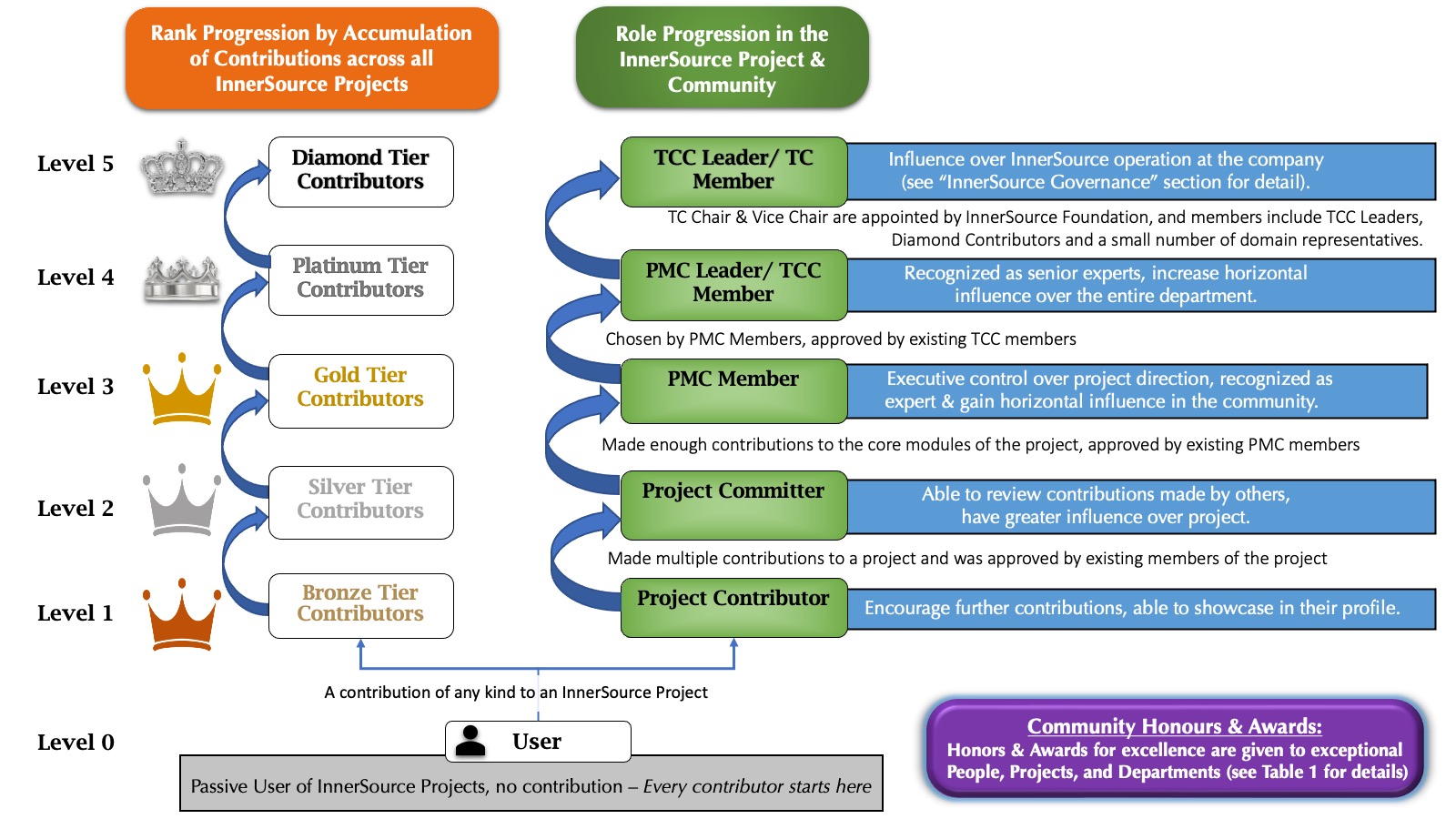}
    \caption{InnerSource Role Progression at Huawei}
    \label{fig:role}
\end{figure*}

\begin{table*}[!htb]
\caption{InnerSource Honors and Awards at Huawei}
\label{t:awards}
\resizebox{\textwidth}{!}{%
\begin{tabular}{@{}p{1.8cm}p{5.5cm}lp{2.5cm}p{5cm}p{1.4cm}@{}}
\toprule
\textbf{Award} &
  \textbf{Selection Criteria} &
  \textbf{Frequency} &
  \textbf{No. of Awards (Monetary Reward - \% of budget for each awardee)} &
  \textbf{Non-monetary Rewards/Benefits} &
  \textbf{Total \% of Budget} \\ \midrule
\multicolumn{2}{l}{\textit{\textbf{Individual Awards}}} &
   &
   &
   &
   \\\midrule
InnerSource Star Award &
  Given to: 1. PMC members/Project maintainers who have demonstrated exceptional leadership or have generated valuable projects and technical knowledge, or 2. Exceptional contributors to InnerSource projects who contributed a significant amount of code, reported/fixed bugs, performed code reviews, and undertaken mentoring responsibilities, or 3. InnerSource ambassadors who have done an outstanding job in promoting InnerSource practices and culture, and attracted new talent. &
  Monthly &
  10 (each receiving 0.25\% of overall budget) &
  The awardees are announced in the InnerSource column and promoted in  Internal newsletters. A star is displayed in their personal profiles, and they are also invited to participate in live broadcasts and discussions. &
  30\% \\\midrule
InnerSource Knight Award &
  Annually given to 10 ``InnerSource Star'' awardees who have brought exceptional value to InnerSource as a whole. &
  Annual &
  10 (each receiving 2.4\% of overall budget) &
  In addition to monetary incentive which is 10 times larger than that given to ``Star'' awardees, ``Knight'' awardees are given \textit{Best Person Memorial Medals}, their names are shared in the InnerSource annual report, and they are invited to participate in the closed InnerSource workshops to discuss further development plans. &
  24\% \\\midrule
\multicolumn{2}{l}{\textit{\textbf{Project Level Awards}}} &
   &
   &
   &
   \\\midrule
   InnerSource Timely Incentive Award &
   Award to projects that achieved significant results based on the four aspects of the InnerSource maturity model (Ref: \url{https://patterns.innersourcecommons.org/p/maturity-model}) - transparency, collaboration, community, and governance. &
   Monthly &
   5 (each project receiving 0.25\% of overall budget) &
   These projects are granted the ``Monthly Active Project'' signpost and logo, and are advertised in corporate-level live broadcasts and workshops. &
   15\% \\\midrule
InnerSource Gold Badge Award &
  Projects are selected based on excellence for InnerSource activities and community building to achieve business value. The award may be given to several projects based on a ranking system - only one project, called the best project of the year, is selected for Rank 1, three projects are selected for Rank 2, and 5 projects are selected for Rank 3. &
  Annual &
  \textit{1 Best Project of Year} (5\% of overall budget), \textit{3 Runner-up projects} (each receiving 4\% of overall budget), \textit{5 Rank 3 projects} (each receiving 1\% of overall budget) &
  The best project is awarded a crystal medal, and all projects are granted customized badges with a ``Best Project" logo. Moreover, the product line president and corresponding management team are introduced to the entire company and are invited to participate in corporate-level live broadcasts and workshops. &
  22\% \\\midrule
\multicolumn{2}{l}{\textit{\textbf{Departmental Awards}}} &
   &
   &
   &
   \\\midrule
InnerSource Black Land Award &
  Special awards for the InnerSource regional operators to create an InnerSource culture in the local research center, generate good InnerSource projects, and encourage more people to participate in InnerSource contributions (by research center region). &
  Annual &
  3 (each receiving 3\% of overall budget) &
  The awardee divisions are given ``InnerSource Black Land Memorial Cup'' and are advertised to the related management levels and are invited to attend the closed workshops to discuss further development. &
  9\% \\ \bottomrule
\end{tabular}%
}
\end{table*}

\subsection{Motivation/Growth Factors}

One of the key features of InnerSource is that it is governed by meritocracy and free choice by participants as to what to work on and when~\cite{InnerSourceAdopting}. The motivations for InnerSource contributors, therefore, are conceptually similar to the higher levels of Maslow's hierarchy of needs, e.g., esteem and self-actualization. In Herzberg’s terminology, these are the `motivation' factors and promote growth both at an individual and community level, ensuring the successful adoption of InnerSource and realization of its numerous benefits, as discussed earlier. There are two main types of such factors - Extrinsic and Intrinsic factors. The extrinsic motivation factors are related to the recognition of an individual's efforts, while the intrinsic motivations could be categorized into Autonomy, Mastery, and Purpose using Pink's Drive terminology~\cite{pink2011drive}. By design, Huawei tried to focus less on monetary rewards and more on reputation-based incentives to avoid the ``Crowding-Out Effect''~\cite{wiersma1992effects}, which suggests that offering too much extrinsic (monetary) reward in the context of pre-existing intrinsic motivation can sometimes undermine the internal motivation for doing that task, thus decreasing the overall performance.

We briefly elaborate on these factors below:

\noindent \textbf{Recognition:} Wanting to be honored and recognized for one's efforts is a basic human instinct and it could act as a significant motivation. Contributing to InnerSource is an opportunity for individuals to showcase their work to a wider audience and gain recognition beyond their individual divisions. This can act as a huge morale booster. Huawei recognizes its importance as well and the incentive program was designed to address this issue.

\noindent \textbf{Autonomy:} In Huawei, the InnerSource working model allows  projects to independently determine and control the project direction over time. The PMCs have autonomy, e.g., in deciding the project direction, incentives within the project, and project working methods. Moreover, individual contributors are also given freedom to choose which projects they want to contribute to and when. This preserves the spontaneity of the InnerSource process and serves to address the need for autonomy.

\noindent \textbf{Mastery:} Mastery refers to the desire to constantly improve one’s skills, which is seen as a primary goal in itself, and not necessarily directly equated with any extrinsic reward. Mastery is also facilitated by autonomy in that InnerSource contributors are able to choose where and how to contribute. This allows them to choose areas of contribution in which they already have expertise, and have a desire to further develop that expertise.

\noindent \textbf{Purpose:} Being a company that is owned by its employees (through an Employee Stock Ownership Program), Huawei recognizes the value of having a sense of loyalty and purpose. Since the InnerSource projects are maintained by the PMCs and contributions are voluntary, developers feel an attachment to the project and strive for its success. This, together with the other motivation factors, gives them a sense of purpose and helps them get closer to achieving self-actualization. 

\section{The Proposed Incentive Program}

With the theoretical foundation for designing an incentive program in place, Huawei has drawn on their extensive experience in designing effective employee appraisal systems to create a structured incentive program for InnerSource. 
The proposed program involves a comprehensive reputation and honor system and a transparent merit-based path for role progression based on an individual's contributions and reputation. There are also special rewards and monetary incentives for exceptional contributors, projects, and divisions, as described below.

\subsection{Contribution-based Reputation System}
Individuals build up their contribution value by making contributions to various InnerSource projects. 
A couple of factors were prominent in designing the reputation scheme -
\begin{enumerate}
    \item  While code contribution to InnerSource projects is counted as the main source of an individual's InnerSource contribution value, other types of contributions, e.g., documentation, bug-reports, code-review, technical discussions are all counted as valid contributions. 
    \item An individual's contribution value is cumulative in nature, i.e. the contribution points and the rewards granted never expire. However, an individual's overall standing in the reputation scale may reduce if others make more contributions. This takes advantage of the loss-aversion instinct of individuals, since no one wants to lose a position they have worked hard to reach, and it keeps them motivated to continue contributing to InnerSource.
\end{enumerate}




\begin{figure*}[!htb]
    \centering
    \includegraphics[width=\textwidth]{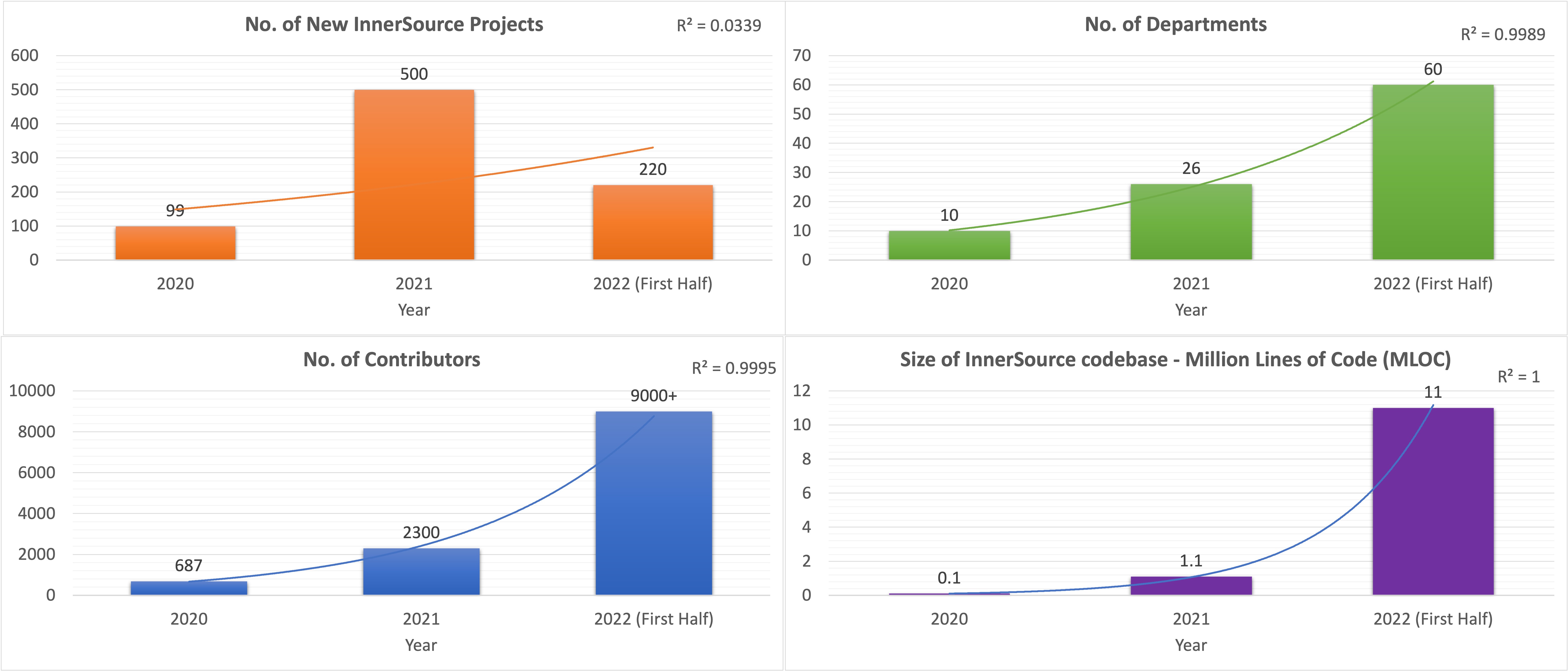}
    \caption{Growth of InnerSource at Huawei}
    \label{fig:trend}
\end{figure*}

\subsection{Role Progression}

In order to promote fairness and transparency, Huawei has adopted merit-based role progression for InnerSource contributors. Contributors can achieve a rank based on  cumulative contribution across all InnerSource projects, as well as achieving a more prominent role in the InnerSource project/community based on their reputation, subject to approval from the community. The proposed role progression ladder for InnerSource contributors at Huawei is shown in Figure~\ref{fig:role}. The required number of contributions needed for rank progression increases exponentially as a contributor's rank increases.

\subsection{Honors and Awards}
As mentioned earlier, Huawei has also introduced InnerSource Awards which are given to exceptional contributors, projects, and departments. There are monetary as well as non-monetary incentives associated with each of the awards. By design, these awards are rare and are only awarded in exceptional cases, so the motivation crowding-out effect does not arise.  The details of the various rewards are described in Table~\ref{t:awards}.
It is worth noting here that while selecting the awardees, Huawei does not simply look at total number of contributions. The individual awards, for example, not only recognize those who made significant technical contributions, but also reward individuals for important managerial and ambassadorial contributions. As for the projects, they are evaluated based on the quality of ancillary resources (e.g., README, documentation, etc.), community management (e.g., issues/PR handling, transparency in governance, etc.), and their usefulness (e.g., number of users, user experience).

\subsection{Wall of Honor} 
The Hall-of-Fame-styled Wall of Honor is also incorporated into Huawei's incentive program to allow more contributions to be recorded and demonstrated over the long term. The homepage of the InnerSource dashboard contains information about  active InnerSource projects as well as other related information about InnerSource, and it is possible to navigate to the ``Wall of Honor''  from the homepage. The Wall of Honor includes the following:

\begin{enumerate}
    \item Honorary presentation of TCs, TCCs, and project PMCs (with personal information, introduction, link to personal page), and  Diamond and Platinum tier contributors.
    \item The annual awards (InnerSource Knight Award, Gold Badge Project Award, Black Land Award) and  awardees for each year.
    \item The monthly awards (InnerSource Project Timely Incentive Item, InnerSource Star) and  awardees for each month.
\end{enumerate}

In addition, an individual's personal profile page can also showcase information regarding their contribution records, awards received, and their current contributor tier and rank. Through this page, visitors can learn about the technical expertise and interests in InnerSource projects, and can initiate mutual technical communication.

\section{Results}

The InnerSource implementation at Huawei is still in its early days - the incentivization program was first proposed in the second half of 2020 and was elaborated and promoted more widely in 2021. As can be seen from Figure~\ref{fig:trend}, the number of departments involved with InnerSource, as well as the number of InnerSource contributors and the total size of the codebase across all InnerSource projects, have increased dramatically since the incentive program was introduced. Although the number of new InnerSource projects in the first half of 2022 is relatively lower, it should be noted that in 2021, a greater proportion of projects started in the second half of the year, and we expect this to be repeated in 2022. Moreover, several projects and contributors have been awarded various awards listed in Table~\ref{t:awards} and some contributors have already reached the ``Gold'' tier (ref: Figure~\ref{fig:role}).

It is not possible at present to determine complete cause and effect for all aspects of the incentive program -- this would require a randomized controlled trial, which is very difficult to conduct at this scale. However, given that several Huawei engineers had cited the lack of incentives as a barrier when InnerSource was re-introduced in late 2019, and that several expressed a positive attitude towards the proposed incentive program since it was introduced, we are inclined to believe the incentive program has had a positive effect.


\section{Conclusion}
As mentioned earlier, numerous organizations across multiple domains have sought to implement InnerSource. However, implementing InnerSource is extremely complex and many organizations fail \cite{capraro2016inner, InnerSourceAdopting}. An appropriate incentivization mechanism is key to its successful implementation. This is no trivial undertaking as both intrinsic and extrinsic motivations need to be incorporated carefully in order to avoid \textit{crowding out} and ensuring that the incentives are appropriate at individual, department and project levels.

The incentive program discussed here has been a key mechanism in ensuring the success of InnerSource at Huawei. We believe this will serve as a good example of how to design an incentive program for InnerSource and be useful to other practitioners in the field.

\bibliographystyle{acm}
\bibliography{references}

\section*{Author biographies}
\begin{wrapfigure}{l}{20mm} 
    \includegraphics[width=1in,height=1in,clip,keepaspectratio]{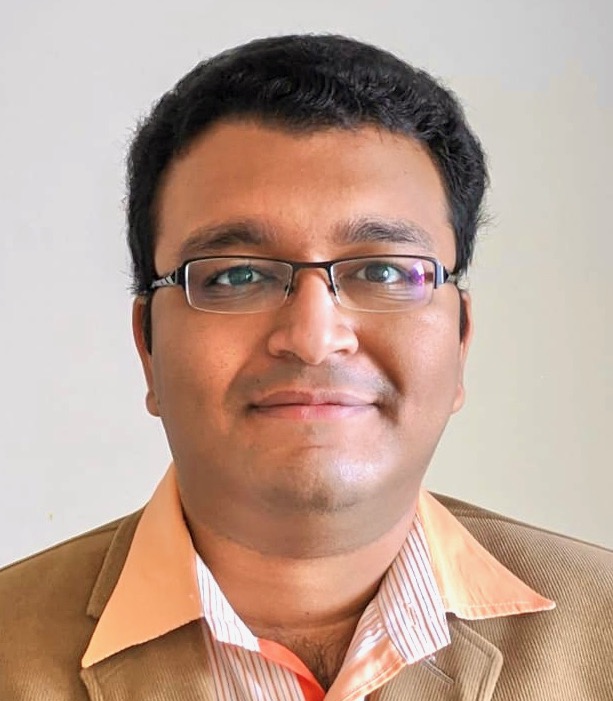}
  \end{wrapfigure}\par
  \textbf{TAPAJIT DEY} is a Post-Doctoral Researcher at Lero – the Irish Software Research Centre at the University of Limerick. He received his Ph.D. from The University of Tennessee. His research interests include open source software, InnerSource, and empirical software engineering. You can find more about him at \url{https://tapjdey.github.io/}. Contact him at: \mbox{tapajit.dey@lero.ie}.
  \par
\vspace{10pt}
\begin{wrapfigure}{l}{25mm} 
    \includegraphics[width=1in,height=1in,clip,keepaspectratio]{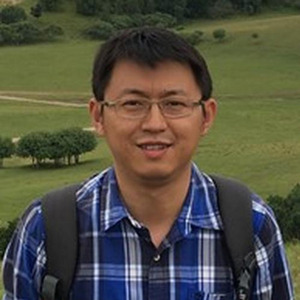}
  \end{wrapfigure}\par
  \textbf{WILLEM JIANG}  is the technical expert of Huawei, a member of the Apache Software Foundation, initiator of Apache Local Community of Beijing. Willem has a passion for mentoring engineers who want to participate in the Open Source Community. Contact him at: \mbox{jiangning9@huawei.com}.
  \par
\vspace{10pt}
\begin{wrapfigure}{l}{25mm} 
    \includegraphics[width=1in,height=1in,clip,keepaspectratio]{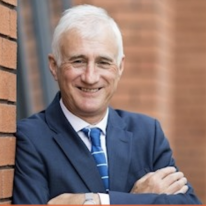}
  \end{wrapfigure}\par
  \textbf{BRIAN FITZGERALD} is Director of Lero – the Irish Software Research Centre and holds the Frederick
Krehbiel Chair in Innovation in Business and Technology at
the University of Limerick. He was also elected President of the Association for Information Systems in 2019. Fitzgerald holds a Ph.D. in Computer Science
from the University of London. Contact him at: bf@lero.ie.
  \par

\end{document}